\documentclass[11pt,tbtags,reqno]{amsart}

\usepackage{pstricks,graphicx}
\usepackage{tabmac}
\usepackage{pb-diagram}
\usepackage{mathrsfs}
\usepackage{latexsym}
\usepackage{amsthm,amsopn,amsfonts,amssymb,epsfig,stmaryrd}
\usepackage[active]{srcltx}

\numberwithin{equation}{section}

\makeatletter
\renewcommand{\subsubsection}{\@startsection
{subsubsection}
{3}
{0mm}
{\baselineskip}
{-0.5\baselineskip}
{\normalfont\normalsize\bfseries}}
\makeatother

\newtheorem{theorem}{Theorem}

\newtheorem{conjecture}[theorem]{Conjecture}

\theoremstyle{remark}

\newtheorem*{acknow}{Acknowledgments}

\def\la{{\lambda}}

\def\cal L{{\mathcal L}}

\def\aa{\alpha}

\def\cd{\circledast}
\def\deg{ {\rm {deg}}}

\newcommand{\tcercle}[1]{\ensuremath{\setlength{\unitlength}{1ex}\begin{picture}(2.8,2.8)\put(1.4,1.4){\circle{2.8}\makebox(-5.6,0){#1}}\end{picture}}}

\def\B{{\mathcal B}}
\def\S{{\mathcal S}}

\let\la\lambda
\let\La\Lambda
\let\Om\Omega

\let\ta\theta

\let\rw\rightarrow

\newcommand{\LL}{\ensuremath{\langle\!\langle}}
\newcommand{\RR}{\ensuremath{\rangle\!\rangle}}

\def\cd{{\circledast}}
\def\Sp{\mathrm{{E}}}

\def\lrw{\leftrightarrow}

\begin{document}

\title[Macdonald polynomials in superspace]{Macdonald polynomials 
in superspace:
conjectural definition and positivity conjectures}

\author{O. Blondeau-Fournier}
\address{D\'epartement de physique, de g\'enie physique et
d'optique, Universit\'e Laval,  Qu\'ebec, Canada,  G1V 0A6.}
\email{olivier.b-fournier.1@ulaval.ca}
\author{P. Desrosiers}

\address{Instituto de Matem\'atica y F\'{\i}sica, Universidad de
Talca, Casilla 747, Talca, Chile.}
\email{patrick@inst-mat.utalca.cl}

\author{L. Lapointe}
\address{Instituto de Matem\'atica y F\'{\i}sica, Universidad de
Talca, Casilla 747, Talca, Chile.}
\email{lapointe@inst-mat.utalca.cl }
\author{P. Mathieu}
\address{D\'epartement de physique, de g\'enie physique et
d'optique, Universit\'e Laval,  Qu\'ebec, Canada,  G1V 0A6.}
\email{pmathieu@phy.ulaval.ca}

   \begin{abstract} We introduce a  conjectural construction
for an extension to superspace of the Macdonald polynomials.
The construction, which depends on certain orthogonality and triangularity
relations, is tested for high degrees.  We conjecture a simple form
for the norm of the Macdonald polynomials in superspace, and a rather 
non-trivial expression for their evaluation.   We study the limiting cases $q=0$ and
$q=\infty$, which lead to two families of Hall-Littlewood
polynomials in superspace.  We also find that the Macdonald polynomials in superspace evaluated at $q=t=0$ or $q=t=\infty$ seem to 
generalize naturally the Schur functions.  
In particular, 
their expansion coefficients in the corresponding Hall-Littlewood bases
appear to be polynomials in $t$ with nonnegative integer coefficients.
More strikingly, we formulate a generalization of the Macdonald positivity conjecture to superspace:  the expansion coefficients of 
the Macdonald superpolynomials expanded into a modified version of 
the Schur superpolynomial basis (the $q=t=0$ family) are polynomials 
in $q$ and $t$ with nonnegative integer
coefficients. 
\end{abstract}


 \maketitle

\section{Introduction}

The Macdonald polynomials \cite{Mac}
can be defined by: (1) their orthogonality with respect to a certain scalar product, 
depending on two parameters $q,t$, and (2) their triangular decomposition in the monomial basis with respect to the dominance ordering. 
The aim of this work is to study the possible extension of the Macdonald polynomials  to superspace along this precise constructive procedure, namely by enforcing triangularity and orthogonality.  In the Jack polynomial 
limit ($q=t^{\alpha}$ and $t\rightarrow 1$), we recall that such an 
extension to superspace was constructed in \cite{DLMcmp2,DLMadv}.
It is important to note that while in the Jack's case 
the construction could 
rely on the supersymmetric version of the 
trigonometric Calogero-Moser-Sutherland model \cite{SS,DLMnpb},
the supersymmetric 
version of the Macdonald operators and the Ruijsenaars-Schneider Hamiltonian 
are still elusive.

  We introduce in Section 2 some preliminary results concerning the Macdonald polynomials and formulate the strategy to be followed in order to investigate their generalization 
without the support of an eigenvalue problem. 
The notion of superspace and 
the relevant properties of the corresponding extension of the Jack polynomials are briefly reviewed in Section 3.
The stage is then prepared for the exploratory investigation of the Macdonald polynomials in superspace (also called Macdonald superpolynomials) 
presented in Section 4:
a previously conjectured form of the scalar product
 \cite[Conj. 34]{DLMadv} is invalidated (no solution to the problem exist),
but by introducing a slight deformation of this scalar product  we end 
up with a new setup for the construction of the Macdonald polynomials 
in superspace. 
The correctness of this construction
is substantiated here by the computation 
of many non-trivial examples which solve (for high degrees) the over-determined systems of equations resulting from the imposed orthogonality 
and triangularity. 
Moreover, we present in the last subsections many conjectures that 
provide further evidence of the
validity of the construction.  In Section~\ref{2con}, we give conjectures for
the norm, the integral version and the evaluation of the 
Macdonald superpolynomials. In Section~\ref{Slimit} we study 
various
limiting cases of the Macdonald superpolynomials.  The limiting cases $q=0$ and
$q=\infty$, which lead to two distinct families of Hall-Littlewood
polynomials in superspace, are worth pointing out.
We also find that the Macdonald polynomials in superspace evaluated at $q=t=0$ or $q=t=\infty$ seem to 
provide natural generalizations of the Schur functions.  In particular, 
we conjecture in Section~\ref{Spositivity} that their 
expansion coefficients in the corresponding Hall-Littlewood bases
are polynomials in $t$ with nonnegative integer coefficients.
Section~\ref{Spositivity} culminates with the most striking 
result of the article:
the extension of the Macdonald positivity conjecture to superspace. 
Concluding observations are made in Section~\ref{Sconc}, and 
finally, tables of Macdonald superpolynomials and Kostka 
coefficients are given at the
end of the article.
  
 \section{Macdonald polynomials}
A partition $\lambda=(\la_1,\la_2,\ldots)$ of degree $d$
is a vector of non-negative integers such that $\lambda_i \geq \lambda_{i+1}$ for all $i$ and such that $\sum_{i} \lambda_i=d$.
The Macdonald polynomials $P_\la=P_\la(x;q,t)$, in the variables 
$x=x_1,x_2,\dots,x_N$, are characterized by the two conditions
 \begin{equation}\label{co12}\begin{array}{lll} 1)& P_{\lambda} =
m_{\lambda} + \text{lower terms},\\
&\\
2)&\LL  P_{\la}| P_{\mu} \RR_{q,t} =0\quad\text{if}\quad \la\ne\mu \, ,
\end{array}\end{equation}
where $N$ is assumed to be larger than the degree of $\lambda$.
The triangular decomposition is given 
with respect  to the usual dominance order on partitions:
\begin{equation}\label{ordre}
   \mu \leq \la\quad\text{ iff }\quad |\mu|=|\la|\quad\text{
and }\quad \mu_1 + \cdots + \mu_i \leq \lambda_1 + \cdots + \lambda_i\quad \forall i \, , \end{equation}
where $|\la|$ stands for the degree of the partition.  The monomial 
functions $m_\la$, in which the triangular expansion is formulated, 
are given by 
 \begin{equation}m_\la=  
{\sum_{\sigma\in S_N}}' x_{\sigma(1)}^{\la_1}\cdots x_{\sigma(N)}^{\la_N},\end{equation}
where the prime indicates a sum over distinct permutations.
Note that we consider in the equation that
$\la_{\ell+1}=\cdots=\la_N=0$ if the partition $\lambda$ is of length 
$\ell=\ell(\la)$ (the 
number of its non-zero parts). 
The  orthogonality relation is defined in the power-sum basis
 \begin{equation} p_\la=p_{\la_1}\cdots p_{\la_\ell}\quad\text{with} \quad p_r=\sum_{i=1}^N x^r_i,
 \end{equation}
by 
\begin{equation}\label{ortoqt}  \LL {p_\la}|
{p_\mu}\RR_{q,t}= z_\la\, \delta_{\la,\mu}\,
\prod_{i=1}^{\ell(\la)}\frac{1-q^{\la_i}}{1-t^{\la_i}}\
,\end{equation}
where
\begin{equation}  \label{zlam}
z_{\la}=\prod_{i \geq 1} i^{n_{\la}(i)} {n_{\la}(i)!}\, ,
\end{equation}
$n_{\la}(i)$  is the number of parts in $\la$ equal to $i$. 

 The Macdonald polynomials can be constructed,
following characterization \eqref{co12},
using the Gram-Schmidt orthogonalization process (choosing any total 
ordering compatible with the dominance ordering).  
We should emphasize that imposing the triangularity 
with respect to the (partial) dominance order diminishes the number of unknowns
in the Gram-Schmidt process and thus
leads to over-determined systems of equations (see Section~2.1). 
Therefore, the question of existence needs to be settled separately. The standard procedure is to show that these polynomials are eigenfunctions of some operators, the Macdonald 
operators in the present case. The existence of non-trivial solutions to an 
overdetermined system is often the hallmark  of an underlying integrable 
model. This is indeed the case here: the operators are actually related to  the Ruijsenaars-Schneider Hamiltonian \cite{Ruij}
(see also for instance \cite[Sect. 1]{vanD} and \cite[Sect. 7.6.2]{RuijR}).

\subsection{Exploring generalizations of the Macdonald polynomials: a simple strategy}
\label{Sstra}
The problem we consider here  is to explore the existence of a particular extension of the  Macdonald polynomials (to be formulated precisely below) in the 
absence of an underlying differential operator with respect to which the 
sought-after generalized polynomials would be eigenfunctions.
The previous remark concerning the over-determinate nature of the corresponding system points toward a simple exploratory strategy  
that we now spell out plainly. 

First, we stress that the consistency of the construction of families of
polynomials characterized by orthogonality and triangularity 
is verified separately for each value of the degree of the partitions.
It is clear that for low values of the degree, where the ordering is total, there will always exist a solution: there is an equal number of unknowns and equations. 
Stated differently (cf. \cite[p. 9]{Maco}), the Gram-Schmidt orthogonalization process ensures a unique solution to the conditions (1) in (\ref{co12}) and a weaker form of (2), namely: 
 \begin{equation}\LL  P_{\la}| P_{\mu} \RR_{q,t} =0\quad\text{if}\quad \la>\mu.\end{equation}

However, for the first value $n$ of the degree at which not all 
partitions of $n$ are comparable in
the dominance ordering, one encounters an over-determined system. 
Then, the critical test amounts to verify the   orthogonality of the polynomials labeled by partitions that cannot be compared. The verification of these orthogonality relations provides a strong hint for the existence of these polynomials under construction. On the other hand, if it fails, this suffices to prove 
that the characterization by orthogonality and triangularity is not correct.
Moreover, one expects this failure to show up for pairs of non-comparable partitions of lowest degree.

\begin{figure}[ht]
\caption{{\footnotesize  Hasse diagrams of the dominance ordering on partitions of degree $6$ and on superpartitions of degree $(4 | 1)$. 
The (super)partitions that can be compared are related by arrows pointing
towards the lowest (super)partitions.}}

\vskip-1.2cm
\label{Posets}
\begin{center}
\begin{pspicture}(0,-1)(12,10.5)
{
\psset{yunit=1 cm,xunit=1 cm,linewidth=1pt}

\psset{linestyle=solid}
\psline{->}(3,0.8)(3,0.45)
\psline{-}(3,0.2)(3,0.5)
\psline{->}(3,1.8)(3,1.45)
\psline{-}(3,1.2)(3,1.5)
\psline{->}(3,6.8)(3,6.45)
\psline{-}(3,6.2)(3,6.5)
\psline{->}(3,7.8)(3,7.45)
\psline{-}(3,7.2)(3,7.5)
\psline{->}(2.2,4.8)(2.5,4.5)
\psline{-}(2.47,4.53)(2.8,4.2)
\psline{->}(2.2,2.8)(2.5,2.5)
\psline{-}(2.47,2.53)(2.8,2.2)
\psline{->}(3.2,3.8)(3.5,3.5)
\psline{-}(3.47,3.53)(3.8,3.2)
\psline{->}(3.2,5.8)(3.5,5.5)
\psline{-}(3.47,5.53)(3.8,5.2)
\psline{->}(2.8,5.8)(2.5,5.5)
\psline{-}(2.53,5.53)(2.2,5.2)
\psline{->}(2.8,3.8)(2.5,3.5)
\psline{-}(2.53,3.53)(2.2,3.2)
\psline{->}(3.8,4.8)(3.5,4.5)
\psline{-}(3.53,4.53)(3.2,4.2)
\psline{->}(3.8,2.8)(3.5,2.5)
\psline{-}(3.53,2.53)(3.2,2.2)
\rput(3,0){{\scriptsize $(1,1,1,1,1,1)$}}
\rput(3,1){{\scriptsize $(2,1,1,1,1)$}}
\rput(3,2){{\scriptsize $(2,2,1,1)$}}
\rput(2,3){{\scriptsize $(3,1,1,1)$}}
\rput(4,3){{\scriptsize $(2,2,2)$}}
\rput(3,4){{\scriptsize $(3,2,1)$}}
\rput(2,5){{\scriptsize $(3,3)$}}
\rput(4,5){{\scriptsize $(4,1,1)$}}
\rput(3,6){{\scriptsize $(4,2)$}}
\rput(3,7){{\scriptsize $(5,1)$}}
\rput(3,8){{\scriptsize $(6)$}}

\psline{->}(9,0.8)(9,0.45)
\psline{-}(9,0.2)(9,0.5)

\psline{->}(8,4.8)(8,4.45)
\psline{-}(8,4.2)(8,4.5)

\psline{->}(10,4.8)(10,4.45)
\psline{-}(10,4.2)(10,4.5)

\psline{->}(9,1.8)(9,1.45)
\psline{-}(9,1.2)(9,1.5)
\psline{->}(9,2.8)(9,2.45)
\psline{-}(9,2.2)(9,2.5)
\psline{->}(9,6.8)(9,6.45)
\psline{-}(9,6.2)(9,6.5)
\psline{->}(9,7.8)(9,7.45)
\psline{-}(9,7.2)(9,7.5)
\psline{->}(9,8.8)(9,8.45)
\psline{-}(9,8.2)(9,8.5)
\psline{->}(9.8,3.8)(9.5,3.5)
\psline{-}(9.53,3.53)(9.2,3.2)
\psline{->}(9.3,4.8)(9,4.5)
\psline{-}(9.03,4.53)(8.7,4.2)
\psline{->}(8.8,5.8)(8.5,5.5)
\psline{-}(8.53,5.53)(8.2,5.2)
\psline{->}(8.2,3.8)(8.5,3.5)
\psline{-}(8.47,3.53)(8.8,3.2)
\psline{->}(9.2,5.8)(9.5,5.5)
\psline{-}(9.47,5.53)(9.8,5.2)
\psline{->}(8.7,4.8)(9,4.5)
\psline{-}(8.97,4.53)(9.3,4.2)
\rput(9,0){{\scriptsize $(0;1,1,1,1)$}}
\rput(9,1){{\scriptsize $(1;1,1,1)$}}
\rput(9,2){{\scriptsize $(0;2,1,1)$}}
\rput(9,3){{\scriptsize $(1;2,1)$}}
\rput(8,4){{\scriptsize $(2;1,1)$}}
\rput(10,4){{\scriptsize $(0;2,2)$}}
\rput(8,5){{\scriptsize $(2;2)$}}
\rput(10,5){{\scriptsize $(0;3,1)$}}
\rput(9,6){{\scriptsize $(1;3)$}}
\rput(9,7){{\scriptsize $(3;1)$}}
\rput(9,8){{\scriptsize $(0;4)$}}
\rput(9,9){{\scriptsize $(4;\,)$}}

\rput(3,-0.8){{\scriptsize\bf(a)}}
\rput(9,-0.8){{\scriptsize\bf(b)}}
}
\end{pspicture}
\end{center}
\end{figure}
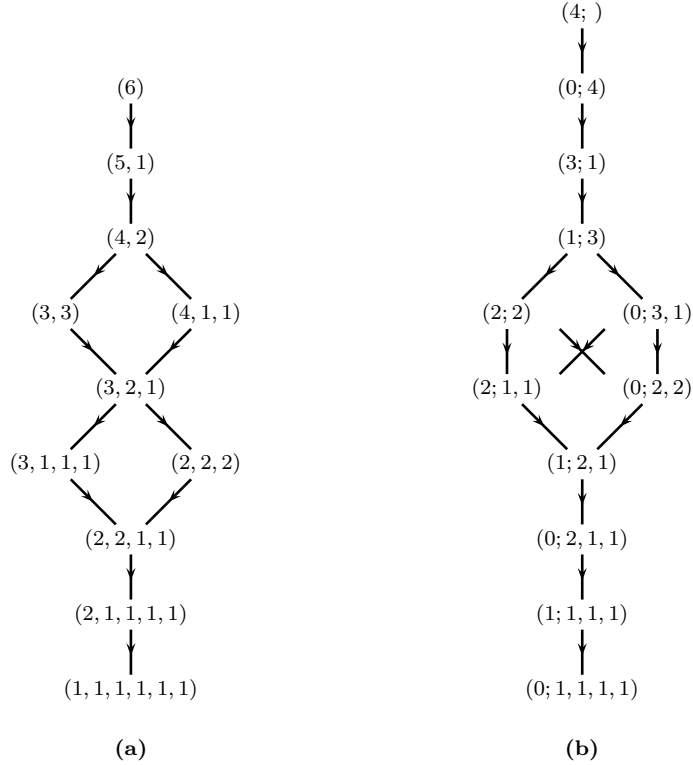

Let us illustrate this simple methodology for the ordinary Macdonald polynomials. The lowest degree at which the dominance ordering is  not total is $n=6$. Indeed, two pairs of partitions cannot be compared: $(3,1,1,1)$ vs $(2,2,2)$ and  $(3,3)$ vs $(4,1,1)$ (cf. Fig. \ref{Posets}(a)). Let us consider the construction of $P_{(3,1,1,1)}$ and $P_{(2,2,2)}$. Denote $(3,1,1,1)$ and $(2,2,2)$ by
$\la^{(1)}$ and $\la^{(2)}$ respectively. Clearly, we have
 \begin{equation}\la^{(i)}\geq (2,2,1,1)\geq (2,1,1,1,1)\geq (1,1,1,1,1,1),\quad i=1,2.\end{equation}
One thus constructs successively $P_{(1,1,1,1,1,1)}$,  $P_{(2,1,1,1,1)}$, $P_{(2,2,1,1)}$, and then, separately, each $P_{\la^{(i)}}$, which are both of the form:
 \begin{equation}P_{\la^{(i)}}= m_{\la^{(i)}}+a^{(i)}_1m_{(2,2,1,1)}+a^{(i)}_2m_{(2,1,1,1,1)}+a^{(i)}_3m_{(1,1,1,1,1,1)},\end{equation}
with given coefficients $a_j^{(i)}$ (depending on  $q$ and $t$) 
obtained by imposing: 
 \begin{equation}\LL  P_{\la^{(i)}}| P_{\mu} \RR_{q,t} =0\quad\text{for}\quad \mu\in\{(2,2,1,1), (2,1,1,1,1), (1,1,1,1,1,1)\}.\end{equation}
The consistency test is to verify the orthogonality relation for the two polynomials labeled by non-comparable partitions, namely: 
 \begin{equation}\LL  P_{\la^{(1)}}| P_{\la^{(2)}} \RR_{q,t} =0\end{equation}
which is known to be satisfied in the present case.

\section{Symmetric polynomials in superspace}

Superspace refers to a pairing of each coordinate $x_i$ with an anticommuting partner $\ta_i$, where $\ta_i^2=0$. A polynomial {in} superspace (also called a superpolynomial) is
a polynomial in the usual (commuting) $N$ variables $x_1,\cdots ,x_N$  and the $N$ Grassmaniann variables $\ta_1,\cdots,\ta_N$. It is  said to be symmetric if it remains invariant with respect to the simultaneous interchange of $x_i\lrw x_j$ and $\ta_i\lrw\ta_j$ for any $i,j$ 
\cite{DLMnpb}.
In other words, introducing for any $\sigma \in S_N$,
\begin{equation}
\mathcal{K}_{\sigma}=\kappa_{\sigma}K_{\sigma},
\qquad\text{where}\quad\begin{cases}
&K_{\sigma}\,:\, (x_1,\dots,x_N) \mapsto (x_{\sigma(1)}, \dots,
x_{\sigma(N)})
\\
&\kappa_{\sigma}\,\;:\, (\theta_1,\dots,\theta_N) \mapsto (\theta_{\sigma(1)}, \dots,
\theta_{\sigma(N)})
,\end{cases}\end{equation}
then a polynomial in superspace $P(x;\theta)$, with $x=(x_1,\ldots,x_N)$ and $\theta=(\theta_1,\ldots,\theta_N)$, is symmetric when
\begin{equation}
\mathcal{K}_{\sigma}P(x;\theta)=P(x;\theta) \qquad {\rm for~all~} \sigma \in S_N 
\, .
\end{equation}

Symmetric superpolynomials are labeled by superpartitions \cite{DLMnpb}.
 A superpartition $\La$ is 
a pair of partitions $(\La^a; \La^s)=(\La_{1},\ldots,\La_m;\La_{m+1},\ldots,\La_N)$, 
 where $\La^a$ is a partition with $m$ 
distinct parts (with possibly one of those parts equal to 0),
and $\La^s$ is a (regular) non-increasing partition.\footnote{When 
$m=0$, $\La=(\,;\la)$ can be thought as the ordinary partition $\la$ 
and the semi-column is omitted.} 
A superpartition $\La$ of the above form is said to have degree 
 $(n|m)$, written for short as deg$(\La)=(n|m)$,  when $\sum_i \La_i =n$.
We refer to $m$ and $n$ respectively as the fermionic degree 
and total degree 
of $\La$.
For example, there are seven superpartitions of degree $(3|1)$:
\begin{equation}
(0 ;1,1,1) ,\, (0 ; 2,1), \, (0 ; 3), \,(1 ; 1,1) ,\, (1 ; 2) ,\, (2 ; 1) ,\, (3;\,),
\end{equation}
and twelve at degree $(4|1)$ (listed in Fig. \ref{Posets}(b)).

To every
superpartition $\La$  we can associate a unique partition $\La^*$
obtained by deleting the semicolon and reordering the parts in
non-increasing order.  
A diagrammatic representation of $\La$ is given by 
the Ferrers diagram of $\La^*$ with
circles added at the end of every part of $\La^a$, and ordered in length as if a circle was a half-box. Finally, we denote by $\La^{\circledast}$ the partition whose Ferrers diagram
is obtained from that of $\La$ by replacing circles by boxes.  
For instance, we have:
\begin{equation} \label{exdia}
\La=(3,1,0;5,4,3):\quad {\tableau[scY]{&&&&\\&&&\\&&&\bl\tcercle{}\\&&\\&\bl\tcercle{}\\
    \bl\tcercle{}}} \qquad 
    \La^*:\quad{\tableau[scY]{&&&&\\&&&\\&&\\&&\\ \\}}\qquad
     \La^\cd:\quad{\tableau[scY]{&&&&\\&&&\\&&&\\&&\\&\\ \\}} .
\end{equation}
 Notice that both $\La^*$ and $\La^{\circledast}$ are ordinary partitions, of respective degree $n$ and $n+m$, if deg$(\La)=(n|m)$.

The natural extension of the usual dominance ordering
to superpartitions  is the following \cite{DLMeva}:
\begin{equation} \label{eqorder1}
 \Omega\leq\Lambda \quad \text{iff}\quad
 \deg(\La)=\deg(\Om) ,
 \quad \Omega^* \leq \Lambda^*\quad \text{and}\quad
\Omega^{\circledast} \leq  \Lambda^{\circledast}.
\end{equation}
 We stress that two pairs of ordinary partitions,
$\Omega^*$ vs $\La^*$ and
$\Omega^{\circledast} $ vs $ \La^{\circledast}$, 
are compared
with respect to the usual  dominance ordering \eqref{ordre}. The lowest degree at which this ordering is not total is $(4|1)$ (see Fig. \ref{Posets}(b)). In that case, two pairs of superpartitions cannot be compared: $(2;1,1)$ vs $(0;2,2)$ and $(2;2)$ vs $(0;3,1)$. For instance, for the first pair, we have:
 \begin{align}\label{exnc}
 &\La=(2;1,1),\qquad\La^*=(2,1,1),\qquad\La^\cd=(3,1,1),\nonumber\\
 &\Om=(0;2,2),\qquad\Om^*=(2,2,0),\qquad\Om^\cd=(2,2,1),
 \end{align} so that
\begin{equation}\label{exnc2}\La^*<\Om^*\qquad\text{but}\qquad \La^\cd>\Om^\cd.
\end{equation} 

We now introduce the two relevant classical bases  of 
symmetric polynomials in superspace \cite{DLMjaco}. The first one is the extension of the monomial polynomials defined as
\begin{equation}
m_\La(x;\theta)={\sum_{\sigma \in S_N} }' \mathcal{K}_\sigma \left(\theta_1\cdots\theta_m x_1^{\La_1}\cdots x_N^{\La_N}\right),
\end{equation}
where the sum is over distinct permutations. 
The other one is the generalization of the power-sum 
symmetric
functions 
\begin{equation}\label{spower}
p_\La=\tilde{p}_{\La_1}\cdots\tilde{p}_{\La_m}p_{\La_{m+1}}\cdots p_{\La_\ell},\qquad\text{where}\quad \tilde{p}_k=\sum_i\theta_ix_i^k\qquad\text{and}\qquad p_\ell=
\sum_ix_i^\ell \, ,
\end{equation}  
where $k\geq 0$ and $\ell \geq 1$.

We conclude this review section by recalling the definition of the  precise objects we want to deform, the Jack superpolynomials. We denote them 
as  $P_\La=P^{(\alpha)}_\La(x,\ta)$; they  are characterized by the two conditions \cite{DLMadv}:
 \begin{equation}\label{SJco}\begin{array}{lll} 1)& P_{\Lambda} =
m_{\Lambda} + \text{lower terms},\\
&\\
2)&\LL  P_{\La}| P_{\Om} \RR_{\alpha} =0\quad\text{if}\quad \La\ne\Om,
\end{array}\end{equation}
where the scalar product is defined as\begin{equation} \label{scap} \LL \, 
{p_\La} \, | \, {p_\Om }\, \RR_\alpha=(-1)^{\binom{m}2}\, \alpha^{{\ell}(\La)}\, z_{\La^s}
\delta_{\La,\Om}\,
 \end{equation}
with $z_{\La^s} $ defined in (\ref{zlam}) and $\ell(\La)=m+\ell(\La^s)=
\ell(\La^{\circledast})$.
The existence of such a basis is proved in \cite{DLMadv} using
the supersymmetric extension of the Calogero-Moser-Sutherland
 model.

\section{The quest for Macdonald polynomials in superspace}

We are now in position to explore the existence of the superspace extension of the Macdonald polynomials obtained by a suitable generalization of the 
construction reviewed previously for the limiting cases of the ordinary Macdonald polynomials and the Jack  superpolynomials.

\subsection{Invalidation of the original conjecture}

 We first consider the conjecture in \cite[Conj. 34]{DLMadv} formulated in terms of the conditions (\ref{SJco}) and the dominance ordering (\ref{eqorder1}),  but with the scalar product  (\ref{scap}) replaced by the following
natural $(q,\,t)$ deformation:\footnote{
The conjecture was originally  presented 
 with respect to the ordering \cite[Corr. 7]{DLMjaco}:
\begin{equation*} \label{eqorder2}
 \Omega\leq'\Lambda \quad \text{iff}\quad
 \deg(\La)=\deg(\Om)
 \quad\text{and}\quad \begin{cases}
& \Omega^* < \Lambda^*\quad \text{or}\quad\\&
\Omega^* = \Lambda^*\quad \text{and}\quad 
\Omega^{\circledast} \leq  \Lambda^{\circledast}.\end{cases}
\end{equation*} 
 The ordering $\leq'$ leads to less over-determined systems of equations
in our construction 
than the ordering $\leq$
given that more pairs are comparable in the ordering $\leq'$  than in 
the ordering $\leq$.  Nevertheless we have verified that the conjecture 
also fails using the ordering $\leq'$ (this time at degree $(6|1)$).}
\begin{equation}  \label{qtspro}\LL {p_\La}|
{p_\Om}\RR^{(F)}_{q,t}=(-1)^{\binom{m}{2}}\,   z_\La(q,t)\delta_{\La,\Omega}\,
,\end{equation}
where
\begin{equation}\label{zti}
\tilde z_\La(q,t)
= z_{\La^s} 
 \prod_{i=1}^{\ell(\La)}
\frac{1-q^{\La^\cd_i}}{1-t^{\La^\cd_i}}\, .\, \end{equation} 
This reduces to (\ref{scap}) when $q=t^{\alpha}$ and $t\rightarrow 1$. The rationale motivating this precise form of the conjectured scalar product is rooted in the naturality of the generalized form of the reproducing kernel:
\begin{equation}\label{falsekernel}
\prod_{i,j}\frac{\left(tx_iy_j+t\theta_i\phi_j;q\right)_\infty}{\left(x_iy_j+\theta_i\phi_j;q\right)_\infty}
=\sum_\La (-1)^{\binom{m}2}
\tilde z_\La(q,t)^{-1}p_\La(x,\theta)\,p_\La(y,\phi)\,,
\end{equation}
where $(a;q)_\infty=\prod_{n\geq 0}(1-aq^n)$ 
(for the details of this correspondence, in particular, the way $\tilde z_\La(q,t)$ is read off this relation, we refer to a similar analysis in \cite[Theo. 33]{DLMjaco}).
Indeed, for all the cases studied previously, namely the classical symmetric functions \cite{DLMjaco} and the Jack polynomials \cite{DLMadv}, the superspace extension of the kernel  is obtained by simply replacing $x_iy_j$ by $x_iy_j+\theta_i\phi_j$.

  As explained in Section \ref{Sstra}, displaying a single counter-example suffices to invalidate the conjecture.  
And the simplest such counter-example is expected to arise at the lowest degree   $(n|m)$ where the ordering is not total, which is  $(4|1)$. This is indeed so: the scalar product of $P_{(2;1,1)}$ and $P_{(0;2,2)}$, polynomials whose labeling superpartitions are  not comparable (cf. (\ref{exnc})-(\ref{exnc2})),  is found to be non-zero.\footnote{It vanishes only in the Jack superpolynomial limit $q=t^\alpha$ and $t\rightarrow 1$ and for $q=t$.} 

\subsection{A new conjecture}

Although (\ref{qtspro}) is natural, the correctness  of the Jack-limiting behavior is  not fully constraining. Actually,
there is a   whole family of acceptable deformations of  (\ref{scap}). 
In particular, the form (\ref{scap}) depends rather weakly upon the parts in $\La^a$: it only depends upon their number  $m$ via the factor $ (-1)^{\binom{m}2}\aa^m$. This suggests to investigate the following simple deformation of $\tilde z_\La$ which amounts to replace the portion of the product depending upon $\La^a$ by 
\begin{equation}
 \prod_{i=1}^{m}\frac{1-q^{\La_i+1}}{1-t^{\La_i+1}}
\quad \rw \quad \prod_{i=1}^{m}\frac{1-q^{\La_i+1}}{1-\tau^{\La_i+1}}
\end{equation}
where $\tau$ is a parameter depending on $q$ and $t$ that
needs to be determined. This form can be obtained from the generalized kernel :
\begin{equation}\label{taukernel}
K^{(\tau)}(x,\theta;y,\phi) = \prod_{i,j} \frac{\left(\tau x_iy_j+ \tau \theta_i\phi_j  ; q\right)_\infty \left(t x_iy_j;q\right)_\infty}{\left(x_iy_j+ \theta_i\phi_j;q\right)_\infty \left(\tau x_iy_j;q\right)_\infty}     
\end{equation}
which reduces to \eqref{falsekernel} when $\tau=t$.
Call the corresponding $\tau$-dependent scalar product $\LL \,|\, \RR^{(\tau)}_{q,t}$. To fix $\tau$, we again consider the lowest degree where the dominance ordering is not total, namely $(4|1)$. As already stressed, at this degree two pairs of superpartitions are non-comparable: $(2;1,1)$ vs $(0;2,2)$ and $(2;2)$ vs $(0;3,1)$.
By enforcing the orthogonality of $P_{(2;1,1)} $ and  $P_{(0;2,2)}$, one gets:
$$\LL P_{(2;1,1)} | P_{(0;2,2)} \RR^{(\tau)}_{q,t}=0 \qquad \iff 
\qquad \tau=q^{-1}$$
As a non-trivial compatibility check, we verify that $P_{(2;2)} $ and $ P_{(0;3,1)}$ are also orthogonal for this choice of $\tau$. 

With $\tau=q^{-1}$, the kernel \eqref{taukernel} simplifies drastically as an infinite number of factors cancel.  It reduces to: 
\begin{equation}\label{Kpri}
K^{(1/q)}(x,\ta;y,\phi)=\prod_{i,j}\left(1 - \frac{\theta_i\phi_j}{q-x_iy_j}\right)\frac{\left(tx_iy_j;q\right)_\infty}{\left(x_iy_j;q\right)_\infty} .\, \end{equation}
The combination $q-x_iy_j$ in \eqref{Kpri} is not natural: $q$ usually appears as a scaling factor of the variables. This can easily be cured by  rescaling the anticommuting variable $\ta_i$ 
by a factor $-q$. This modification does not affect the resulting Macdonald polynomials since all their components are multiplied by the same number of anticommuting variables (this alteration only changes their norm). The effect of the replacement $\theta_i\rw -q\theta_i$ in the kernel is to transform it into
\begin{equation}
K(x,\ta;y,\phi)=\prod_{i,j}\left(1 {+}\frac{\theta_i\phi_j}{1-q^{-1}x_iy_j}\right)\frac{\left(tx_iy_j;q\right)_\infty}{\left(x_iy_j;q\right)_\infty}\,=K_0(x;y)\prod_{i,j}\left(1 {+}\frac{\theta_i\phi_j}{1-q^{-1}x_iy_j}\right) .\, \end{equation}
where $K_0(x;y)$ is the  kernel in the 
absence of anticommuting variables. (The last expression is to be compared with \cite[eq. (5.14)]{DLMadv} 
in the Jack case.) 
With
\begin{equation}\label{Ksp}
K(x,\ta;y,\phi)=\sum_\La  (-1)^{\binom{m}2}{z_\La(q,t)}^{-1} \, p_\La(x,\ta)\,p_\La(y,\phi),\end{equation}
our new scalar product reads: 
\begin{equation}\label{newsp}
\LL {p_\La}|{p_\Om}\RR_{q,t}=(-1)^{\binom{m}2}\,z_\La(q,t)\delta_{\La,\Omega},
\end{equation}
where
\begin{equation}\label{newz}
 z_\La(q,t)
= z_{\La^s} 
\, q^{|\La^a|} \prod_{i=1}^{\ell(\La^s)}
\frac{1-q^{\La^s_i}}{1-t^{\La^s_i}}\, ,\, \end{equation}
indeed, a rather simple form.
We note however that this does not precisely reduces to the expression
$ \LL \, 
{p_\La} \, | \, {p_\Om }\, \RR_\alpha$ written in (\ref{scap}), but rather 
to $\alpha^{-m}\, \LL \,  {p_\La} \, | \, {p_\Om }\, \RR_\alpha$.
However, such a modification does not affect 
the resulting Jack superpolynomials since only polynomials with the 
same fermionic degree $m$ are compared.

It is thus natural to conjecture the existence of a superspace generalization of the Macdonald polynomials, written $P_\La=P_\La(x,\theta;q,t)$, as follows:

\begin{conjecture}\label{con1}
For each superpartition $\La$ there is a unique symmetric superpolynomial  $P_\La$ such that: 
\begin{equation}\label{mac1}
\begin{array}{lll} 1)& P_{\Lambda} =
m_{\Lambda} + \text{lower terms},\\
&\\
2)&\LL  P_{\La}| P_{\Om} \RR_{q,t} =0\quad\text{if}\quad \La\ne\Om.
\end{array}\end{equation}
where lower terms are considered with respect to 
the ordering (\ref{eqorder1}) and the scalar product is 
defined in (\ref{newsp}) and (\ref{newz}).   
\end{conjecture}

This conjecture has been heavily tested. The cases that have been checked are listed in Table~\ref{tab1}. 
Examples of Macdonald polynomials in superspace are presented in Table~\ref{tab2}.

\begin{table}[ht]
\caption{The different degrees that have been tested at which the system is over-determined. }
\label{tab1}
\footnotesize{\begin{center}
\begin{tabular}{|c|c|c|c|c|c|c|c|c|c|} 
\hline
\hline
&&&&&&&&&
\\
 degree $(n|m)$& $(4|1)$ & $(5|2)$ & $(5|1)$ & $(6|3)$ &$(6|2)$ & $(6|1)$   & $(7|3)$  & $(8|3)$  & $(9|4)$ 
 \\&&&&&&&&&\\
$\#\,\text{equations}$&66&136&171& 45 &378& 435& 171&528&45\\
&&&&&&&&&\\ 
$\#\,\text{unknowns}$&64&128&163& 43 &351&401&163&482&43\\&&&&&&&&&\\
\hline\hline

\end{tabular}
\end{center}
}
\end{table}

 From the examples in Table~\ref{tab2}, 
we observe that a property of the Macdonald polynomials that 
does not hold  in superspace is the invariance with respect to $(q,t)\mapsto(q^{-1},t^{-1})$: in general, whenever $m>0$ we have that 
  \begin{equation}\label{nonsymprop}
  P_\La(x,\ta;q,t)\ne P_\La(x,\ta;q^{-1},t^{-1}) 
  \end{equation}
(cf. \cite{Mac} property VI (4.14)--(iv) for $m=0$). This is already manifest from the simplest displayed case $P_{(1;0)}$. In fact, if we set $\bar P_\La(x,\ta;q,t)=P_\La(x,\ta;q^{-1},t^{-1})$, then  Conjecture \ref{con1} is equivalent to saying that $\bar P_\La(x,\ta;q,t)$ is the unique polynomial satisfying 1) and 2),  but with the scalar product replaced by \begin{equation}\label{newspt}
\LL {p_\La}|{p_\Om}\RR^{'}_{q,t}=(-1)^{\binom{m}2}\,\bar z_\La(q,t)\delta_{\La,\Omega},\qquad  {\rm where} \quad \bar z_\La(q,t)
= z_{\La^s} 
\, t^{-|\La^a|} \prod_{i=1}^{\ell(\La^s)}
\frac{1-q^{\La^s_i}}{1-t^{\La^s_i}}\, .\end{equation}
As will be explained below,  this new scalar product is more convenient in certain circumstances since it allows to take the limit $q\to 0$.    

\begin{table}[ht]
\caption{  Sample of small degree Macdonald polynomials in superspace expanded in the monomial basis (excluding those containing a single monomial term).} 
\label{tab2}
\small{ 
\begin{center}
\begin{tabular}{|c|l|} 
\hline
& \\
deg($\La$) & $P_\La $ \\ & \\ \hline & \\
$(1|1)$ 
  &   $P_{(1;\,)}=m_{(1;\,)}+\frac{q(1-t)}{1-qt} \,m_{(0;1)}$ \\
  & \\ \hline
&
\\
$(2|1)$ 
 & $P_{(1;1)}=m_{(1;1)}+\frac{q(1-t^2)}{1-qt^2}  \,m_{(0;1,1)}$
\\ 
 & $P_{(0;2)}=m_{(0;2)}+\frac{1-t}{1-qt}  \,m_{(1;1)}+\frac{(1+q)(1-t)}{1-qt}  \,m_{(0;1,1)}$
\\ 
 & $P_{(2;\,)}=m_{(2;\,)}+\frac{q^2(1-t)}{1-q^2t}  \,m_{(0;2)}+\frac{q(1+q)(1-t)}{1-q^2t}  \,m_{(1;1)}+\frac{q^2(1+q)(1-t)^2}{(1-qt)(1-q^2t)} \,m_{(0;1,1)}$
\\ & \\ \hline
 & \\
$(2|2)$
 & $P_{(2,0;\,)}=m_{(2,0;\,)}+\frac{q(1-t)}{1-qt}  \,m_{(1,0;1)}$
 \\ & \\ \hline &\\
$(3|2)$ &$P_{(1,0;2)} =m_{(1, 0;2)}+\frac{(1+qt)(1-t)}{1-qt^2} \,m_{(1, 0;1, 1)}$\\
&$P_{(2,0;1)}=m_{(2,0;1)}+\frac{q(1-t)}{1-qt} \,m_{(1,0;2)}+\frac{2q(1-t)} {1-qt}   \,m_{(1,0;1,1)}$\\
&$P_{(2,1;)}= m_{(2,1;)}+\frac{q(1-t)}{1-qt} \,m_{(2,0;1)} -\frac{q^2t(1-q)(1-t)}{(1+qt)(1-qt)^2}  \,m_{(1,0;2)}+\frac{q^2(1-t)^2}{(1-qt)^2}   \,m_{(1,0;1,1)}$\\
& $P_{(3,0;)}= m_{(3,0;)}+\frac{q(1-t)}{1-q^2t} \,m_{(2,1;0)} +\frac{q(1+q)(1-t)}{1-q^2t } \,m_{(2,0;1)} +\frac{q^2(1-t)}{ 1-q^2t} \,m_{(1,0;2)}$
\\
&$\qquad \qquad 
+\frac{q^2(1+q)(1-t)^2} {(1-qt)(1-q^2t) }  \,m_{(1,0;1,1)}$
\\ & \\ \hline &\\
$(4|1)$   &$P_{(1;1,1,1)}=m_{(1;1,1,1)}+\frac{q(1-t^4)}{1-qt^4}m_{(0;1,1,1,1)}$ \\
& $P_{(0;2,1,1)}=m_{(0;2,1,1)}+\frac{1-t^3}{1-qt^3}m_{(1;1,1,1)}+\frac{(1-t)(3qt^2+t^2+2t+2qt+q+3)}{1-qt^3}m_{(0;1,1,1,1)}$   \\

& $P_{(1;2,1)}=m_{(1;2,1)}+\frac{q(1-t^2)}{1-qt^2}m_{(0;2,1,1)}+ \frac{(1-t)(t+2qt+q+2)}{1-qt^2}m_{(1;1,1,1)}$  \\
& $\qquad \qquad + \frac{q(1-t^2)(1-t)(3qt^2+t^2+2t+2qt+q+3)}{(1-qt^3)(1-qt^2)}m_{(0;1,1,1,1)}$ \\

&$P_{(2;1,1)}=m_{(2;1,1)}+\frac{q(1-t)}{1-qt}m_{(1;2,1)}+\frac{q^2(1-t)(1-qt^3)}{(1-q^2t^3)(1-qt)}m_{(0;2,1,1)}$ \\
&$ \qquad\qquad + \frac{q(1-t)(3q^2t^3+qt^3-q-3)}{(1-q^2t^3)(1-qt)}m_{(1;1,1,1)}+\frac{q^2(1-t)^2(3qt^2+t^2+2t+2qt+q+3)}{(1-q^2t^3)(1-qt)}m_{(0;1,1,1,1)}$ \\

& $P_{(0;2,2)} =m_{(0;2,2)}+\frac{1-t}{1-qt}m_{(1;2,1)}+\frac{(1+q)(1-t)}{1-qt}m_{(0;2,1,1)} $ \\
& $ \qquad \qquad + \frac{(1-t)^2(t+2qt+q+2)}{(1-qt^2)(1-qt)}m_{(1;1,1,1)}+\frac{(1+q)(1-t)^2(t+2qt+q+2)}{(1-qt^2)(1-qt)}m_{(0;1,1,1,1)}$ \\

&$P_{(2;2)}=m_{(2;2)}+\frac{q^2(1-t^2)}{1-q^2t^2}m_{(0;2,2)}+\frac{(1+q)(1-t)}{1-qt}m_{(2;1,1)}+ \frac{q(1+q)(1-t)(1-qt^2)}{(1-q^2t^2)(1-qt)}m_{(1;2,1)}$ \\
&$\qquad \qquad + \frac{q^2(1+q)(1-t)(1-t^2)}{(1-q^2t^2)(1-qt)}m_{(0;2,1,1)}+ \frac{q(1+q)(1-t)^2(t+2qt+q+2)}{(1-q^2t^2)(1-qt)}m_{(1;1,1,1)}$ \\
& $\qquad \qquad + \frac{q^2(1+q)(1-t)^2(1-t^2)(t+2qt+q+2)}{(1-qt^2)(1-q^2t^2)(1-qt)}m_{(0;1,1,1,1)}$ \\

&$P_{(0;3,1)}=m_{(0;3,1)}+\frac{(1+q)(1-t)}{1-qt}m_{(0;2,2)}+\frac{1-t^2}{1-q^2t^2}m_{(2;1,1)}+ \frac{(1+q)(1-t)(1-qt^2)}{(1-q^2t^2)(1-qt)}m_{(1;2,1)}$ \\
&$\qquad \qquad + \frac{(1-t)(-2q^2t^2-qt^2-2q^3t^2-qt+q^2t+2q+q^2+2)}{(1-q^2t^2)(1-qt)}m_{(0;2,1,1)}+ \frac{(1+q)(1-t)^2(t+2qt+q+2)}{(1-q^2t^2)(1-qt)}m_{(1;1,1,1)}$ \\
& $\qquad \qquad + \frac{(1+q)(1-t)^2(3q^2t+q^2+2qt+2q+t+3)}{(1-q^2t^2)(1-qt)}m_{(0;1,1,1,1)}$  \\

&$\qquad  \cdots$ 

\\ & \\ \hline
\end{tabular}
\end{center}
}
\end{table}

\def\B{{\mathcal B}}
\subsection{The norm, the integral version and the specialization}
\label{2con}

For a box $s=(i,j)\in\La$ (i.e., in row $i$ and column $j$), introduce the following arm-lengths and leg-lengths:
\begin{align} 
a(s)= \La^*_i-j , \qquad\qquad& \,\tilde a(s)=  \La^\cd_i-j,\nonumber\\
l(s)= ({\La}^*)'_j-i,\qquad\qquad & \tilde l(s)= ({\La}^\cd)'_j-i,
\end{align}
where $({\La}^*)'$ and $({\La}^\cd)'$ stand for the 
conjugate of the partitions $\La^*$ and $\La^{\cd}$ respectively.
We now define
\begin{equation}w_\La (q,t) =  \prod_{ s \in \B(\La) }  (1 - q^{ a(s)+1 } t^{ \tilde{l}(s) }  ) \, ,
\end{equation}
where $\B(\La)$ denotes the set of boxes in the diagram of  $\La$ that do not
appear at the same time in a row containing a circle {and} in a
column containing a circle (this excludes for instance the boxes $(3,1),\,(3,2)$ and $(5,1)$ of $\La$ whose diagram is found in \eqref{exdia}). 
These are the required ingredients for formulating our conjecture for the norm expression.
\begin{conjecture}\label{con2}
Let $P_\La$ be the superpolynomial of Conjecture \ref{con1} and its norm be defined by
\begin{equation} \|P_\La\|^2= (-1)^{\binom{m}{2}}\LL {P_\La}|{P_\La}\RR_{q,t}.
\end{equation}
Then, the norm is given by
\begin{equation}
\|P_\La\|^2
 =q^{ | \La^a | }  \frac{w_\La (q,t)}{ w_{\La' }(t,q)} = q^{ | \La^a | } \prod_{ s \in \B(\La) } \frac{1 - q^{ a(s)+1 } t^{ \tilde{l}(s) }}{1 - q^{ \tilde{a}(s) } t^{ l(s)+1 }},
\end{equation} where 
 $\La'$ is the conjugate of $\La$ 
(the superpartition whose diagram is obtained by interchanging the 
rows and the columns in the diagram of $\La$). 
\end{conjecture}

 This expression for the combinatorial norm  generalizes naturally both that of the Macdonald polynomials \cite[Eq. (VI.6.19)]{Mac} and that of the Jack superpolynomials established in  \cite{LLN,DLMeva}.
For example, the above formula yields
$$\|P_{(3,0;1)}\|^2 =\frac{q^3 (1-q)^2 ( 1+q)}{(1-t)(1-q^2 t)}\qquad\text{and}\qquad \|P_{(2,1,0;)}\|^2 =q^3.
$$

Given our conjectured norm expression, it is natural to guess the form of  the proper integral version of the Macdonald superpolynomials:
\begin{conjecture}\label{con3}
The  monomial expansion coefficients of the following normalization of the 
Macdonald superpolynomials
\begin{equation} \label{eqintform}
J_\La = w_{\La'}(t,q) P_\La.
\end{equation} are polynomial in $q,\,t$ with integral coefficients (albeit not necessarily positive).
\end{conjecture}

This is a natural generalization of the integral form of the usual Macdonald polynomials \cite[Eq. (VI.8.3)]{Mac}
and the $(q,\,t)$-version of \cite[Conj. 33]{DLMadv}.  The extension to superspace of
the Macdonald positivity conjecture (see Conjecture~\ref{ConjMac}) will be
given in terms of $J_\La $.

Finally, we turn to the conjectured expression for the specialization.
Let $F(x;\theta)$ be a polynomial in superspace of fermionic degree $m$
and suppose that $N\geq m$.
The evaluation of  $F(x;\theta)$ is defined as
\begin{equation}\label{EqDefSpecialI}
\Sp_{N,m}[F(x;\ta)] := \left[
\frac{\partial_{\theta_m} \cdots \partial_{\theta_1}
F(x;\theta)}{V_m(x)}\right]_{\substack{x_1=u_1,\cdots,x_N=u_N}},\end{equation}
where the specialization values $u_i$ are given by
\begin{equation}
u_i =\frac{t^{i-1}}{q^{\text{max}(m-i,0)}}
\end{equation}
 and where
\begin{equation}
V_m(x)={\prod_{1\leq i<j\leq m}(x_i-x_j)}
\end{equation}
is the Vandermonde determinant in the variables $x_1,\dots,x_m$. 

The following conjecture involves a combinatorial  number $\zeta_\La$ that has no analogue for ordinary partitions. We consider the partial filling of the squares of $\La$ defined as follows: in each fermionic square of $\La$ write the number of bosonic squares above it; $\zeta_\La$ is obtained by adding up these numbers. Here are three examples for which it is non-vanishing:
\begin{equation}
\zeta_{(1,0;2,2)}=2:\quad {\tableau[scY]{&\\&\\2&\bl\tcercle{} \\\bl\tcercle{} \\ }}  \qquad \zeta_{(3,1,0;2)}=1:\quad {\tableau[scY]{0&0&&\bl\tcercle{}\\& \\1&\bl\tcercle{} \\\bl\tcercle{} \\ }}  \qquad
\zeta_{(2,1,0;3,1)}=3 :\quad {\tableau[scY]{&&  \\1&1&\bl\tcercle{}  \\1&\bl\tcercle{}\\ \\ \bl\tcercle{} }}\end{equation}
We also need to introduce, 
for $\Lambda$ a superpartition of fermionic degree $m$,
the skew diagram $\S \Lambda=
\Lambda^{\circledast}/\delta^{(m+1)}$, where $\delta^{(m)}$ is the staircase
partition $(m-1,m-2,\dots,1,0)$,
and the quantity $b(\lambda)=\sum_{i} (i-1)\lambda_i$, where $\lambda$
is a partition.
Note that $b(\lambda/\mu)$ will stand for $b(\lambda)-b(\mu)$. 
\begin{conjecture}\label{con4}
Let $\La$ be of fermionic degree $m$ and suppose that 
$N\geq \ell(\La^{\circledast})$.
Then the evaluation formula for the Macdonald superpolynomials $J_\La$ defined in Conjecture~\ref{con3} is:
\begin{equation}
\Sp_{N,m}[J_\La(x,\ta;q,t)]= \frac{{t^{\zeta_\La}}\, t^{b(\S \Lambda)}}
{q^{(m-1) |\Lambda^a/\delta^{(m)}| - b(\Lambda^a/\delta^{(m)})}} {\prod_{(i,j)\in \S\La}\left(1-q^{(j-1)}t^{N-(i-1)}\right)} 
\end{equation}
\end{conjecture}

For $m=0$, this reduces to the usual evaluation formula for the Macdonald polynomials \cite[Eq. (VI.6.11')]{Mac}, while for $q=t^\aa$ and $t\rw1$, it reduces to that of the Jack superpolynomials \cite[Theo. 26]{DLMeva}.

To illustrate this formula, consider the evaluation of $J_{(3,1,0;1,1)}$. 
 Here $m=3$ and we set $N=7$, so that $(x_1,\cdots,x_7)=(q^{-2},tq^{-1},t^2,t^3,t^4,t^5,t^6)$. 
We have
 \begin{equation}
(m-1)|\La^a/\delta^{(m)}|-b(\La^a/\delta^{(m)})=2\cdot 1-0=2\, ,   
\quad 
 b(\S\Lambda)=b(42111/321)=11-4=7
.\end{equation}
In the following product the contributing boxes are $(1,4),\, (4,1),\, (5,1)$:
    \begin{equation}\prod_{(i,j)\in \S(\La)}\left(1-q^{(j-1)}t^{N-(i-1)}\right)=(1-q^3t^7)(1-t^4)(1-t^3).\end{equation}
Finally, since  $\zeta_{(3,1,0;1,1)}=0$, we have 
 \begin{equation}\Sp_{7,3}[J_{(3,1,0;1,1)}]=q^{-2}t^7(1-t^3)(1-t^4)(1-q^3t^7).\end{equation}

\subsection{Limiting cases} \label{Slimit}
\begin{figure}[ht]
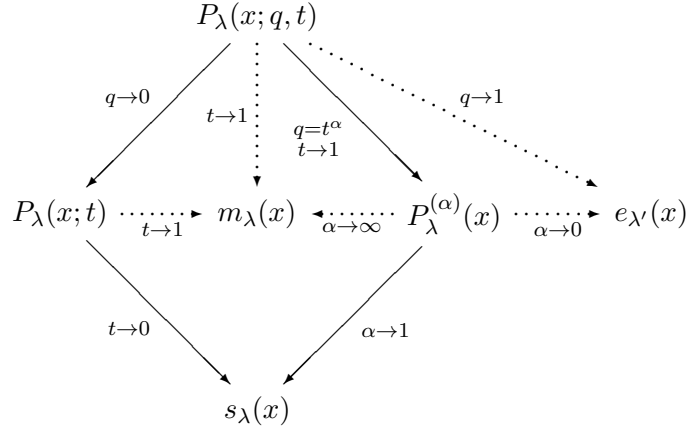

\caption{{\footnotesize Classical limiting  cases.  The dotted arrows indicate limits not related to the Schur functions}}
\label{climits}
$$\qquad \qquad \qquad
\dgARROWLENGTH=2.8em
\begin{diagram}
\node{}\node{P_\la(x;q,t)} \arrow{sw,t}{q\to 0}\arrow{s,l,..}{t\to 1} \arrow{se,b}{\substack{q=t^\alpha\\t\to 1}}\arrow{ese,t,..}{q\to 1} \node{}\node{}\\
\node{P_\la(x;t)} \arrow{se,b}{t\to 0}\arrow{e,b,..}{t\to 1}\node{m_\la(x)} \node{P^{(\alpha)}_\la(x)} \arrow{w,b,..}{\alpha\to \infty}\arrow{sw,b}{\alpha\to 1} \arrow{e,b,..}{\alpha\to 0} \node{e_{\la'}(x)}  \\
\node{}  \node{s_\lambda(x)}  \node{}\node{}
\end{diagram}
\qquad \qquad 
$$
\end{figure}

We recall that for any superpartition $\La=(;\lambda)$ of fermionic degree $m=0$,  the superpolynomial  $P_\La(x,\ta;q,t)$  reduces to the standard Macdonald polynomial $P_{\lambda}(x;q,t)$, whose distinct limiting cases  \cite{Mac} are summarized in Fig.~\ref{climits}.   We now aim at generalizing this picture
to superspace.

The simplest cases are those directly related to the Jack polynomials.  Indeed, as previously mentioned,   with $q=t^\alpha$ 
and $t\rightarrow 1$ in $P_\La(x,\ta;q,t)$ , 
 the monic Jack superpolynomial $P^{(\alpha)}_\La(x,\ta)$ is recovered. Moreover, it has been previously established  \cite{DLMadv} that  $P^{(\alpha)}_\La(x,\ta)$ tends to $m_\La(x,\ta)$ as $\alpha\to \infty$  while it tends to $(-1)^{\binom{m}2} e_{\La'}(x, \ta)$ as $\alpha \to 0$, where the elementary superpolynomials are defined as
\begin{equation}
e_\La (x, \ta)=\tilde{e}_{\La^a_1}(x, \ta)\cdots\tilde{e}_{\La^a_m}(x, \ta)\,e_{\La^s_{1}}(x)\cdots e_{\La^s_\ell}(x),\end{equation}
with
\begin{equation}
\tilde{e}_n(x, \ta)= m_{(0;1^n)}(x,\ta) \qquad\text{and}\qquad 
e_n(x)=m_{(1^n)}(x).\end{equation} 
Similarly, for all tested Macdonald superpolynomials, we found
 \begin{equation}
P_{\La}(x,\ta;q,1)=m_\La(x, \ta)\qquad\text{and}\qquad P_{\La}(x,\ta;1,t)=(-1)^{\binom{m}2} e_{\La'}(x, \ta).
\end{equation}
This means that each dotted limit of Fig.\ \ref{climits} has a simple generalization to superspace.

Consider now the   Hall-Littlewood limit.  In the standard case, $P_\lambda(x;t)$ is equal to $P_\La(x;q,t)$ whenever $q=0$, so it is natural to define the Hall-Littlewood superpolynomial as 
\begin{equation}P_\La(x,\ta;t)=P_\La(x,\ta;0,t)\,.\end{equation} 
 From Table \ref{tab2},  we  see that  many superpolynomials $P_\La$ simply reduce to $m_\La$ when $q=0$.
This extreme simplicity is, however, an artifact of the smallness of the  degree of these examples. 
Another noteworthy feature is that the scalar product \eqref{newsp} is degenerate when $q=0$.
Consequently, our construction based on the extension of the reproducing 
kernel does not provide a natural  scalar product  for the  $P_\La(x,\ta;t)$'s.
In particular, their existence would not follow from
a proof of the existence of the Macdonald polynomials in superspace.

There is however another possible generalization of the Hall-Littlewood polynomial to superspace.  Indeed, the standard Macdonald polynomials are invariant under the transformation $(q,t)\mapsto (q^{-1},t^{-1})$,  which implies that   $P_\lambda(x;t^{-1})=P_\la(x;\infty,t)$.  This suggests to define $\bar P_\La(x,\ta;t^{-1})=P_\La(x,\ta;\infty,t)$,  which is equivalent to defining
\begin{equation}\bar P_\La(x,\ta;t)=\bar P_\La(x,\ta;0,t)\,,\end{equation}   
 where $\bar P_\La(x,\ta;q,t) =P_\La(x,\ta;q^{-1},t^{-1})$.   As could be
expected from the non-symmetry in \eqref{nonsymprop},
one gets  (see Table~\ref{tab2}) that the families  
$\bar P_\La(x,\ta;t)$ and   $P_\La(x,\ta;t)$ are distinct.
The advantage of the family $\bar P_\La$ is that it 
inherits a scalar product from \eqref{newspt}.  
\begin{conjecture}\label{con5}
The Hall-Littlewood superpolynomial $\bar P_\La(x,\ta;t)=P_\La(x,\ta;\infty,t^{-1})$ is the unique superpolynomial satisfying 1) and 2) of Conjecture \ref{con1} but with the second condition replaced by
\begin{equation}
\LL \bar P_\Om , \bar P_\La \RR'_{0,t} =0,\qquad \Om\neq \La.
\end{equation}
where the scalar product is that given in \eqref{newspt} when $q=0$.  
\end{conjecture}

\begin{figure}[ht]
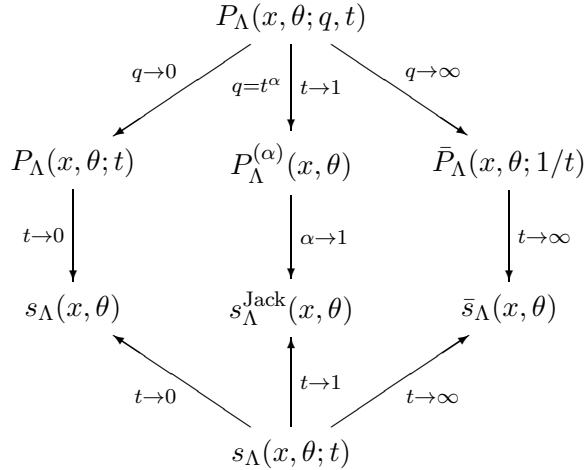

\caption{{\footnotesize Limiting  cases in superspace}}

\label{newlimits}
$$\begin{diagram}
\node{}\node{P_\La(x,\theta;q,t)} \arrow{sw,t}{q\to 0}\arrow{s,lr}{q=t^\alpha}{t\to 1}\arrow{se,t}{q\to \infty} \node{}\\
\node{P_\La(x,\theta;t)} \arrow{s,l}{t\to 0} \node{P^{(\alpha)}_\La(x,\theta)} \arrow{s,r}{\alpha\to 1} \node{\bar P_\La(x,\theta;1/t)} \arrow{s,r}{t\to \infty} \\
\node{s_\La(x,\theta)}  \node{s^\mathrm{Jack}_\La(x,\theta)}   \node{\bar s_\La(x,\theta)}  \\
\node{} \node{s_\La(x,\theta;t)}\arrow{nw,b}{t\to 0}\arrow{n,r}{t\to 1}\arrow{ne,b}{t\to \infty} \node{}
\end{diagram}
$$
\end{figure}

It is tempting to conclude from the above conjecture that the appropriate Hall-Littlewood superpolynomials should be  the $\bar P_\La $, 
but as we will see shortly, both families enjoy remarkable and somewhat entangled properties.  A similar separation phenomenon occurs for the Schur 
superpolynomials.  We know from Fig. \ref{climits} that the standard Schur polynomial $s_\la$ can be seen either as the Hall-Littlewood polynomial 
$P_\la(x;0)$ or as the Jack polynomial $P^{(1)}_\la(x)$.   The generalization of the limiting cases  to superspace is given in Fig.~\ref{newlimits} and provides us with three possible Schur superpolynomials that do not depend upon any extra 
parameter:
\begin{equation*}s_\Lambda(x,\theta)=P_\Lambda(x,\ta;0,0),\qquad s^\mathrm{Jack}_\Lambda(x,\theta)=P_\Lambda(x,\ta;1,1), \qquad \bar s_\Lambda(x,\theta)=P_\Lambda(x,\ta;\infty,\infty),\end{equation*}
These superpolynomials  
are respectively equal to the limits $t\to 0$, $t\to 1$,  $t\to \infty$ of   
\begin{equation} s_\La(x,\theta;t)=P_\La(x,\ta;t,t).\end{equation}
We stress that when $\La=(;\la)$ has fermionic degree $m=0$, the latter reduces to $P_{\la}(x;t,t)$, which is independent of $t$ and equal to the standard Schur polynomial $s_{\la}(x)$.   According to Conjecture~\ref{con1}, the one-parameter Schur polynomials $s_\La(x,\theta;t)$ should be the unique triangular monic superpolynomials that are also  orthogonal with respect to the following scalar product: \begin{equation}
\LL p_\La | p_\Om \RR_{t,t}=(-1)^{\binom{m}2}     z_{\La^s} t^{|\La^a |} \, \delta_{\La,\Om}.
\end{equation}
Note that the latter equation is well defined for all $t\neq 0,\infty$.  From a combinatorial point of view however, the positivity properties described below confirm that  the cases $t=0,\infty$ are the most interesting even though the
scalar product is degenerate in those cases.

\subsection{A Macdonald positivity conjecture in superspace}
\label{Spositivity}

As was just mentioned, by setting $\alpha=1$ in the Jack superpolynomial case, one obtains a candidate superspace version of the Schur polynomials.\footnote{We stress  that these are completely different from the  
so-called supersymmetric Schur polynomials (see e.g.,
\cite{Stem}) which have no relation with anticommuting variables. For more details, see \cite[Remark 14]{DLMjaco}.} However, these superpolynomials do not seem to have distinguished combinatorial properties such as positivity and integrality.
This is neatly illustrated by the following example: 
\begin{multline}
\label{laid}
s^\mathrm{Jack}_{(1,0;3)}=\lim_{t\rw 1}P_{(1,0;3)}(x,\ta;t,t)\\=m_{(1,0;3)}+\frac12m_{(2,0;2)}+\frac78m_{(1,0;2,1)}+\frac14m_{(2,0;1,1)}-\frac18m_{(2,1;1)}+\frac34m_{(1,0;1,1,1)}.
\end{multline}
By contrast, the limits $t\to0, \infty$ depicted in Fig.\ \ref{newlimits} respectively lead to  
\begin{equation}
\label{beau}s_{(1,0;3)}=m_{(1,0;3)}+m_{(2,0;2)}+m_{(1,0;2,1)} +m_{(2,0;1,1)}
+m_{(1,0;1,1,1)},
\end{equation}
\begin{equation}
\label{beau2}\bar s_{(1,0;3)}=m_{(1,0;3)}+m_{(1,0;2,1)} 
+m_{(1,0;1,1,1)}.
\end{equation}
 From the consideration  of all the examples we could construct (cf. Table \ref{tab1}),  
 we formulate the following conjecture.
 \begin{conjecture}
 The expansion coefficients $K_{\La \Om}$ and $\bar K_{\La \Om}$ defined by
 \begin{equation}
s_\La =\sum_{\Om\leq \La} \bar K_{\La \Om}\, m_{\Om}\qquad \text{and}
\qquad \bar s_\La =\sum_{\Om\leq \La}  K_{\La \Om}\, m_{\Om}
 \end{equation}
are non-negative integers.
\end{conjecture}
It is interesting and somewhat surprising to see that in superspace, the Schur polynomials with positive coefficients  cannot be reached from the Jack side but solely from the Hall-Littlewood ones.    
Even more remarkably, 
it appears that the positivity of the change of basis matrix between the Schur and Hall-Littlewood polynomials generalizes to 
superspace.
\begin{conjecture}\label{conjHL}
 The expansion coefficients $K_{\La \Om}(t)$ and $\bar K_{\La \Om}(t)$ defined by
 \begin{equation}
  s_\La =\sum_{\Om\leq \La} \bar K_{\La \Om}(t)\, P_{\Om}(t)\qquad \text{and}\qquad \bar s_\La =\sum_{\Om\leq \La} K_{\La \Om}(t)\, \bar P_{\Om}(1/t)
 \end{equation}
are polynomials in $t$ with nonnegative integer coefficients.  Furthermore,  
when $t=1$ they satisfy $\bar K_{\La \Om}(1)=\bar K_{\La \Om}$ and $K_{\La \Om}(1)=K_{\La \Om}$.
 \end{conjecture}

We end this section by presenting an outstanding conjecture that both supports the correctness of our construction of the Macdonald superpolynomials and the related definitions of the Schur superpolynomials: the superspace version of the Macdonald positivity conjecture (see \cite{Macc} or \cite{Mac}, eq VI (8.18?), and \cite{Hai} for its proof). 
Introduce first the endomorphism $\varphi$ of
$\mathbb Q(q,t)[p_1,p_2,p_3,\dots;\tilde p_0,\tilde p_1,\tilde p_2,\dots]$, 
defined by its action on the power-sums as\footnote{This endomorphism corresponds in the non-fermionic case 
to the operation that sends $f[X]$ to $f[X(1-t)]$ in $\lambda$-ring notation.
This operation is called ``plethysm'' by some authors (see for instance
Section 3.8 of \cite{Ber}).}
\begin{equation}
\varphi(p_n)=(1-t^n)p_n\qquad\text{and}\qquad 
\varphi(\tilde p_n)=\tilde p_n.
 \end{equation}
We then introduce a deformation of the Schur superpolynomials $s_{\La}(x,\theta)$
as
\begin{equation}\label{defM}
 S_\Lambda (x,\ta;t) = \, \varphi \bigl(s_\Lambda (x,\ta)\bigr).
 \end{equation}
\begin{conjecture}\label{ConjMac}
 The coefficients $ K_{\Omega \Lambda} (q,t)$ in the expansion of the
integral form of the Macdonald superpolynomials (see Conjecture~\ref{con3})
\begin{equation}
 J_\Lambda (x,\ta;q,t)  = \sum_\Omega K_{\Omega \Lambda} (q,t) \,S_\Omega(x,\ta;t)
 \end{equation}
are polynomials in $q$ and $t$ with nonnegative  integer coefficients
(tables of  coefficients $ K_{\Omega \Lambda} (q,t)$ can be found at the
end of the article).
\end{conjecture}

Although the Macdonald positivity conjecture in superspace 
is formulated in terms of the family of Schur superpolynomials $s_\Om$,   our computer calculations revealed that the generalized $q,t$-Kostka coefficients $K_{\Omega \Lambda}(q,t)$  are related to the second family $\bar s_\La$ in a very elegant way.
\begin{conjecture} We have
$
K_{\Om\La}= K_{\Omega \Lambda} (0,1)$ and $K_{\Om\La}(t)= K_{\Omega \Lambda} (0,t)$.
\end{conjecture}

\section{Conclusion and outlook}\label{Sconc}

Since the discovery of the Jack superpolynomials \cite{DLMcmp2}, a natural 
question has been to determine their Macdonald counterpart. 
The obvious starting point was to look for the supersymmetric version of the trigonometric Ruijsenaars-Schneider model and define the Macdonald superpolynomials as the common eigenfunctions of all the conservation laws. However, no such candidate Hamiltonian has been found so far: this is still an  open problem. We stress that to demonstrate the existence of the Macdonald superpolynomials 
it is almost imperative to  determine the operators that have them as eigenfunctions.

Under these circumstances, it is natural to proceed by brute force and investigate the existence of  Macdonald polynomials in superspace constructed along a precise procedure (triangularity and orthogonality).
This type of construction leads to over-determined systems of equations
that only in a few remarkable cases have a solution.  The experimental 
confirmation that 
the over-determined systems of equations we obtain have a solution
(cf. Table \ref{tab1}) gives strong evidence that 
we have defined the right 
Macdonald polynomials in superspace. 
We got further confirmation of the correctness of the characterization
 \eqref{mac1}
by examining in Section \ref{2con} three other combinatorial properties, which happen to have the expected form generalizing what was obtained in 
the Jack superpolynomial case. 
 
In addition, we have identified (to be published elsewhere) the vanishing conditions generalizing \cite{FJMM2} (see \cite{FJMM1} in the Jack's case and \cite{DLM0} for the superspace version).
The essential 
modification here is that the admissibility condition $\la_i-\la_{k+i}\geq r$ of  \cite{FJMM2} is changed to $\La_i^\cd-\La^*_{i+k}\geq r$.\footnote{Quite interestingly, such admissible superpartitions for $r=2$
describe the basis of states in the superconformal minimal models $\mathcal{SM}(2,4k+4)$ (see \cite{Mel},\cite[Appendix]{FJM} and the introduction of \cite{DLM0} for more details).} 

The discovery
of combinatorially richer versions of the Schur functions 
in superspace also  
open many avenues.   For instance, new tableaux combinatorics should explain
the monomial positivity of $s_\Lambda$ and $\bar s_{\Lambda}$, and  
a connection with representation theory could hopefully be unraveled.


But what appears to be the strongest validation
of the present construction is clearly the superspace
version of the Macdonald positivity conjecture.
 In a forthcoming publication, we will present
other conjectures related to the $q,t$-Kostka coefficients in
superspace, such as symmetry properties that generalize VI (8.14) and VI (8.15)
of \cite{Mac}.  One of those conjectures states explicitly how the 
$q,t$-Kostka coefficients provide a non-trivial refinement of 
the $q,t$-Kostka coefficients
($q,t$-Kostka coefficients appear to be positive sums of $q,t$-Kostka 
coefficients in superspace).  We can thus hope that the $q,t$-Kostka
coefficients in superspace will help shed some light on the famously difficult 
combinatorics of the $q,t$-Kostka coefficients.

\begin{table}[ht]
\caption{  $K_{\Om \La}(q,t)$ for degree $(1|1)$.} 
\label{tab11}
\begin{center}
\begin{tabular}{c| c |c | } 
 & $(1;\,) $ & $(0;1)$ \\ \hline
$(1;)$ & $1$ & $q$ \\ \hline
$(0;1\,)$ & $t$ & $1$ \\ \hline
\end{tabular}
\end{center}
\end{table}
\begin{table}[ht]
\caption{ $K_{\Om \La}(q,t)$ for  degree $(2|1)$.} 
\label{tab21}
\begin{center}
\begin{tabular}{c| c |c | c | c | } 
 & $(2;\,) $ & $(0;2 )$ & $(1;1 )$ & $(0;1,1 )$ \\ \hline
$(2;\,)$ & $1$ & $q^2$ & $q $ & $q^3 $ \\ \hline
$(0;2)$ & $t$ & $ 1$ & $qt$ & $q$\\ \hline
$(1;1)$ & $t$ & $ qt$ & $1$ & $q$\\ \hline
$(0;1,1)$ & $t^3$ & $ t$ & $t^2$ & $1$\\ \hline
\end{tabular}
\end{center}
\end{table}
\begin{table}[ht]
\caption{$K_{\Om \La}(q,t)$ for   degree $(2|2)$.} 
\label{tab22}
\begin{center}
\begin{tabular}{c| c |c | } 
 & $(2,0;\,) $ & $(1,0;1)$ \\ \hline
$(2,0;\,)$ & $1$ & $q$ \\ \hline
$(1,0;1)$ & $t$ & $1$ \\ \hline
\end{tabular}
\end{center}
\end{table}
\begin{table}[ht]
\caption{ $K_{\Om \La}(q,t)$ for  degree $(3|1)$.} 
\label{tab31}
\begin{center}
\begin{tabular}{c| c |c | c | c | c| c| c|} 
                  & $(3;\,) $ & $(0;3 )$ & $(2;1 )$ & $(1;2 )$ & $(0;2,1 )$ & $(1;1,1 )$ & $(0;1,1,1 )$ \\ \hline
$(3;\,)$ & $1$ & $q^3$ & $q+q^2 $ & $q^2+q^4 $  & $q^4+q^5$ & $q^3$ & $q^6$  \\ \hline
$(0;3)$ & $t$ & $ 1$ & $qt + q^2t$ & $q+q^2t$  & $q+q^2$  & $q^3t$ & $q^3$ \\ \hline
$(2;1)$ & $t$ & $ q^2t$ & $1+qt$ & $q+q^2t$  & $q^2+q^3t$  & $q$ & $q^3$\\ \hline
$(1;2)$ & $t^2$ & $ qt$ & $t+ qt^2$ & $1+q^2 t^2$ & $q+q^2t$  & $qt$  & $q^2$\\ \hline
$(0;2,1)$ & $t^3$ & $ t$ & $t^2+qt^3$ & $t+qt^2$ & $1+qt$  & $qt^2$  & $q$\\ \hline
$(1;1,1)$ & $t^3$ & $ qt^3$ & $t+t^2$ & $t+qt^2$ & $qt+qt^2$  & $1$  & $q$\\ \hline
$(0;1,1,1)$ & $t^6$ & $ t^3$ & $t^4+t^5$ & $t^2+t^4$ & $t+t^2$  & $t^3$  & $1$\\ \hline
\end{tabular}
\end{center}
\end{table}
\begin{table}[ht]
\caption{ $K_{\Om \La}(q,t)$ for  degree $(3|2)$.} 
\label{tab32}
\begin{center}
\begin{tabular}{c| c |c | c | c |c| } 
 & $(3,0;\,) $ & $(2,1;\, )$ & $(2,0;1 )$ & $(1,0;2 )$ & $(1,0;1,1 )$ \\ \hline
$(3,0;\,)$ & $1$ & $q$ & $q+q^2 $ & $q^2 $  & $q^3$ \\ \hline
$(2,1;\,)$ & $qt$ & $ 1$ & $q+q^2t$ & $q^3t$ & $q^2$\\ \hline
$(2,0;1)$ & $t$ & $ t$ & $1+qt$ & $q$ & $q$\\ \hline
$(1,0;2)$ & $t^2$ & $ qt^3$ & $t+qt^2$ & $1$ & $qt$\\ \hline
$(1,0;1,1 )$ & $t^3$ & $ t^2$ & $t+t^2$ & $t$ & $1$\\ \hline
\end{tabular}
\end{center}
\end{table}

\begin{acknow}
We thank the referee for his constructive comments. This work was  supported by NSERC, FQRNT, 
FONDECYT (Fondo Nacional de Desarrollo Cient\'{\i}fico y
Tecnol\'ogico de Chile) grants \#1090016 and \#1090034, and by CONICYT (Comisi\'on Nacional de Investigaci\'on Cient\'ifica y Tecnol\'ogica de Chile) via the Redes De Colaboraci\'on RED4 and the Programa de Investigaci\'on Asociativa ACT56.
\end{acknow}

\end{document}